\documentclass{moriond}

\def\Journal#1#2#3#4{{#1} {\bf #2}, #3 (#4)}

\def\PLB{{\em Phys. Lett.}  B}

\def\be{\begin{equation}}
\def\ee{\end{equation}}
\def\bea{\begin{eqnarray}}
\def\eea{\end{eqnarray}}

\usepackage{xspace}
\newcommand{\Photo}{\includegraphics[height=35mm]{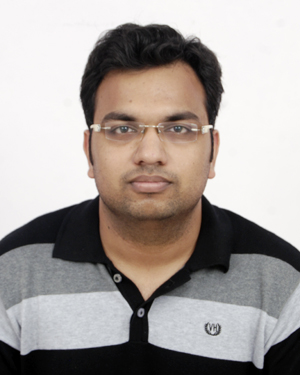}}
\newcommand{\MET}{{$E_{T}^{miss}$}\xspace}
\newcommand{\MT}{{$M_{\rm T}$}\xspace}
\newcommand{\pt}{{$p_{\rm T}$}\xspace}
\newcommand{\mindphijmet}{{$\min \Delta\phi({\rm jet},E_{\rm T}^{miss})$}\xspace}

\newcommand{\mh}{$M_{H}$}

\begin{document}
\vspace*{4cm}
\title{SEARCH FOR A HEAVY HIGGS BOSON IN $\mathrm{H}\rightarrow \mathrm{ZZ}\rightarrow 2\mathrm{\ell}2\mathrm{\nu}$ CHANNEL IN pp COLLISIONS WITH CMS DETECTOR AT THE LHC.}

\author{ ARUN KUMAR }

\address{Department of Physics and Astrophysics, University of Delhi,\\
Delhi-110007, India}

\maketitle\abstracts{
A search for a heavy Higgs boson in the $\mathrm{H}\rightarrow \mathrm{ZZ}\rightarrow 2\mathrm{\ell}2\mathrm{\nu}$ decay channel, where $\mathrm{\ell}$ = $\mathrm{e}$ or $\mathrm{\mu}$, in pp collisions at a center-of-mass energy of 7 and 8 TeV is presented. The search is optimized separately for the vector boson fusion and the gluon fusion production processes. No significant excess is observed above the background expectation. A re-interpretation of the results as a search for an electroweak singlet extension of the Standard Model is also presented.}

\section{Introduction}

On $4^{th}$ July 2012, two major experiments ATLAS and CMS at LHC announced the discovery of a boson at mass around 125 GeV~\cite{2012gk,2012gu} which by far looks compatible with the SM Higgs boson. However, it is not quite sure yet that this discovered boson is the one favored by the EWK fits and whether it is solely responsible for the unitarization of the cross-section for the scattering of longitudinally polarized vector bosons. In addition, there are models, such as general two-Higgsdoublet models or models in which the SM Higgs boson mixes with a heavy electroweak singlet, which predict the existence of additional resonances at high mass, with couplings similar to the SM Higgs boson. This paper reports a search for a SM-like Higgs boson in $\mathrm{ZZ}\rightarrow 2\mathrm{\ell}2\mathrm{\nu}$ decay mode~\cite{13014} in mass range 200 GeV - 1000 GeV. Gluon-gluon fusion and vector boson fusion (VBF) production modes are considered. Along with that, direct search has been carried out for a heavy electroweak singlet Higgs boson with a narrower width than the SM one. This analysis uses full data recorded by the CMS experiment which corresponds to an integrated luminosity of 5.0 ${\mathrm{fb}}^{-1}$ at $\sqrt{s} = $ 7 TeV and 19.7 ${\mathrm{fb}}^{-1}$ at $\sqrt{s} = $ 8 TeV.

\section{Analysis Strategy}\label{sec:analysis_strategy}
In the signal we have two leptons consistent with $\mathrm{Z}$ boson decay along with large transverse missing energy (\MET) coming from another $\mathrm{Z}$ decay to neutrinos. With this topology in mind, we have the following main backgrounds: $\mathrm{Z} + \mathrm{jets}$, $\mathrm{t\bar{t}}+\mathrm{jets}$, $\mathrm{WW}$, $\mathrm{WZ}$, $\mathrm{ZZ}$ and subdominant backgrounds like single-top and $\mathrm{W}+\mathrm{jets}$ etc. Analysis strategy is devised to reduce these backgrounds. The event selection is mainly divided into two parts: \textit{preselection} and \textit{final selection}. Events are first selected with dilepton triggers having \pt thresholds 17 GeV and 8 GeV on the leading and other lepton respectively. Then, preselection starts with the selection of two well identified and isolated leptons of same flavor ($\mathrm{e^{+}e^{-}}$ or $\mathrm{\mu^{+}\mu^{-}}$) with \pt $>$ 20 GeV having invariant mass within a 30 GeV window centered on $\mathrm{Z}$ boson mass. The \pt of the dilepton system is required to be greater than 55 GeV. To reduce the effect of fake-\MET arising from jet mismeasurements, mainly in $\mathrm{Z}+\mathrm{jets}$ process, a selection is applied such that events are removed if the angle in the azimuthal plane between the \MET and the closest jet, \mindphijmet, is smaller than 0.5 radians. The top-quark related backgrounds are suppressed by applying a veto on events having a $\mathrm{b}$ tagged jet by the jet probability algorithm. To further suppress these backgrounds, a veto is applied on events containing a ``soft muon'' with \pt $>$ 3 GeV, which is typically produced in the leptonic decay of a $\mathrm{b}$ quark. An extra lepton veto is also applied to reduce the leptonic decays of $\mathrm{WZ}$ background.
After the preselection events are categorized based on jet multiplicity. Jets are selected with \pt $>$ 30 GeV. First the events are checked for VBF tagging. An event is tagged as VBF like if the two highest \pt jets have a minimal pseudorapidity separation of $\mathrm{|\Delta\eta| >}$ 4 and invariant mass $>$ 500 GeV. Remaining events are categorized into exclusive 0-jet and inclusive 1-jet categories.\\
The Higgs boson search is performed in two ways: by analysing a kinematic shape of the most discriminating variable called, ``shape analysis'', or by counting the events after full selection called, ``cut-and-count'' analysis. The final selection is based on two variables \MET and transverse mass of Higgs (\MT). In the case of VBF category a constant \MET $>$ 90 GeV and no \MT requirements are used. For non-VBF categories, a Higgs mass dependent selection is applied on these variables for cut-and-count analysis which is shown in Table~\ref{tab:met_mtcuts}.
\begin{table}[htp]
\begin{center}
\begin{tabular}{c|ccccc}\hline
\mh (GeV)    & 200        & 300        & 400         & 500        & $\geq$600 \\
\hline
\hline
\MET (GeV)     & $>80$     & $>100$     & $>110$      & $>110$     & $>110$  \\
\MT (GeV)    & $180 - 220$ & $180 - 270$ & $350 - 450$  & $400 - 600$ & $500 - \infty$ \\
\hline
\end{tabular}
\end{center}
\caption{
      Higgs boson mass-dependent selection for \MET and
      \MT variables in the gluon fusion analysis.}
\label{tab:met_mtcuts}
\end{table}
The shape-based analysis is done by fitting the \MT distribution, in the ``0-jet'' and ``$\geq$1-jets'' categories, and the \MET distribution in the VBF category. For the ``0-jet'' and ``$\geq$1-jets'' categories we apply the \MET requirements as shown in Table~\ref{tab:met_mtcuts}.
\subsection{Background Estimation}
The $\mathrm{ZZ}$ and $\mathrm{WZ}$ backgrounds are modeled using Monte Carlo simulation and are normalized to their respective NLO cross sections with $\mathrm{MCFM}$. The remaining backgrounds are estimated from data using data-driven techniques. $\mathrm{Z}+\mathrm{jets}$ is estimated using $\mathrm{\gamma}+\mathrm{jets}$ events in data. This has the advantage of being a large statistics sample, which has similar characteristics to $\mathrm{Z}$ production: i.e. production mechanism, underlying event conditions, pileup scenario and hadronic recoil. $\mathrm{\gamma}+\mathrm{jets}$ events are reweighted as a function of boson \pt in each jet multiplicity bin to match to $\mathrm{Z}+\mathrm{jets}$ in data. The procedure takes into account the dependence of the \MET on the associated hadronic activity. To further reduce the discrepancies due to to pileup difference and photon triggers in $\mathrm{\gamma}+\mathrm{jets}$ and $\mathrm{Z}+\mathrm{jets}$, a reweighting is applied based on number of reconstructed vertices. A mass is assigned to each photon for calculation of \MT based on probability density function constructed from the measured dilepton invariant mass distribution in $\mathrm{Z}+\mathrm{jets}$ events.\\
The background processes that do not involve a $\mathrm{Z}$ resonance are referred to as non-resonant backgrounds which includes $\mathrm{t\bar{t}}+\mathrm{jets}$, $\mathrm{WW}$, $\mathrm{W}+\mathrm{jets}$, single top-quark events and $\mathrm{Z}\rightarrow \mathrm{\tau\tau}$. We estimate the contribution of non-resonant backgrounds by using a control sample of events with dileptons of difference flavor ($\mathrm{e^\pm\mu^\mp}$) that pass the full analysis selection. The non-resonant background in the $\mathrm{e^{+}e^{-}}$ and $\mathrm{\mu^{+}\mu^{-}}$ final states is estimated by applying a scale factor ($\mathrm{\alpha}$) to the selected $\mathrm{e^\pm\mu^\mp}$ events. The scale factor is defined as:
\begin{eqnarray}
N_{\mu\mu} = \alpha_{\mu} \times N_{e\mu}, \qquad
N_{ee} = \alpha_{e} \times N_{e\mu},
\end{eqnarray}
and it is computed from the sidebands (SB) of the $\mathrm{Z}$ peak ($40< M(\ell\ell) <70$ GeV and $100< M(\ell\ell)<200$ GeV by using the following relations:
\begin{eqnarray}
\alpha_{\mu} = \frac{N_{\mu\mu}^\mathrm{SB}}{N_{e\mu}^\mathrm{SB}}, \qquad
\alpha_{e} = \frac{N_{ee}^\mathrm{SB}}{N_{e\mu}^{SB}},
\end{eqnarray}
where $N_{ee}^\mathrm{SB}$, $N_{\mu\mu}^\mathrm{SB}$, and $N_{e\mu}^\mathrm{SB}$ are the number of events
in the $\mathrm{Z}$ sidebands counted in a top-enriched sample of $\mathrm{e^{+}e^{-}}$ and $\mathrm{\mu^{+}\mu^{-}}$, and $\mathrm{e^\pm\mu^\mp}$ final states, respectively. Such samples are selected by requiring \MET$>$70 GeV and a $\mathrm{b}$-tagged jet in the events.
The measured values of $\mathrm{\alpha}$ with the corresponding statistical uncertainties
are $\mathrm{\alpha_{\mu}= 0.71 \pm 0.04}$ and $\mathrm{\alpha_{e} = 0.44 \pm 0.03}$.
\section{Systematic Uncertainties}
Systematic uncertainties include experimental uncertainties on the selection and measurement of the reconstructed objects, theoretical uncertainties on the signal and background processes which are derived from Monte Carlo simulation, and uncertainties on backgrounds determined from control samples in data. Details of the systematics are given in Ref.~\cite{13014}.
\section{Results}
\begin{figure}
\begin{minipage}{0.48\linewidth}
\centerline{\includegraphics[width=0.7\linewidth]{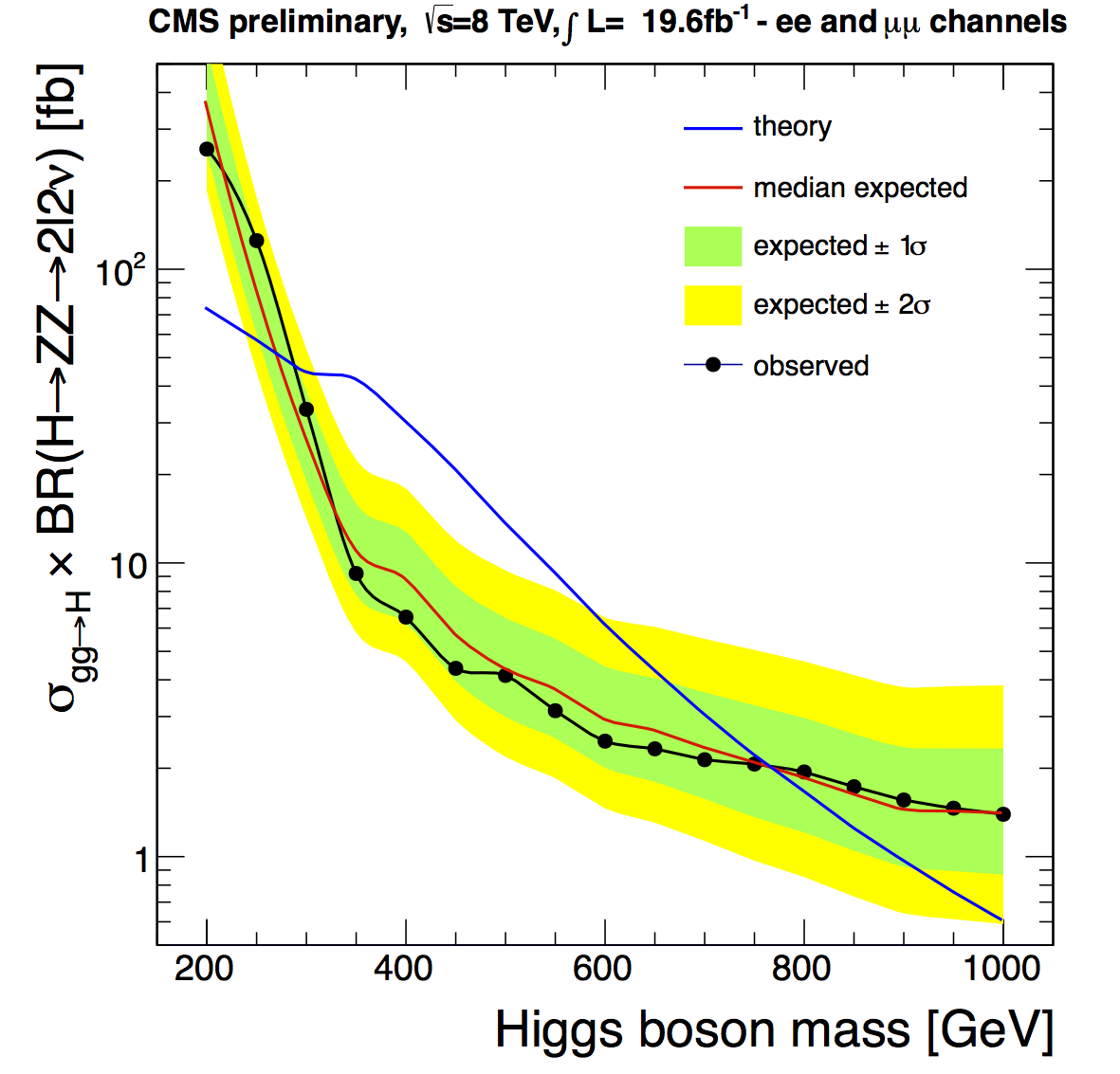}}
\end{minipage}
\hfill
\begin{minipage}{0.48\linewidth}
\centerline{\includegraphics[width=0.7\linewidth]{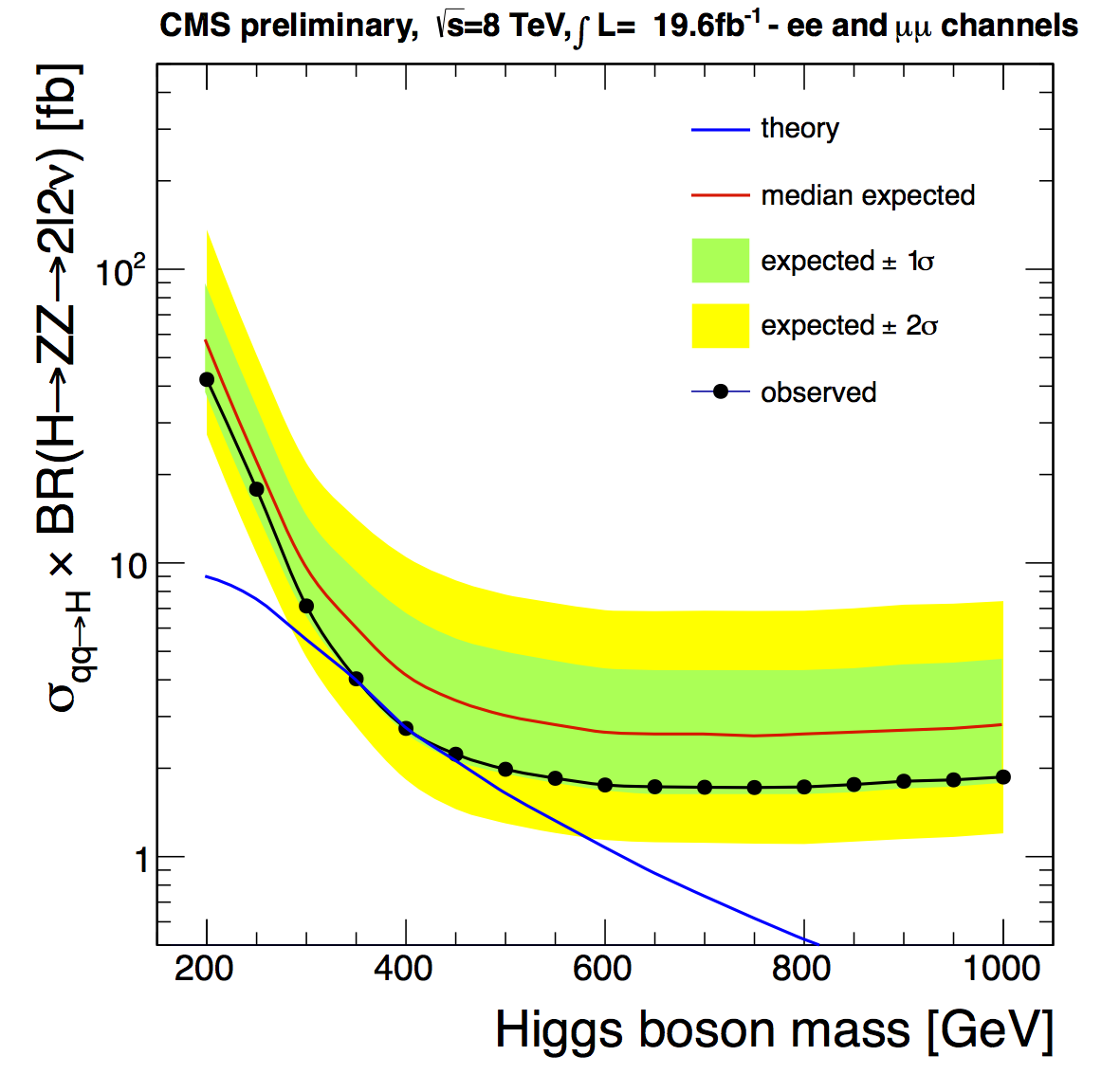}}
\end{minipage}
\caption{Left: Upper limits at 95\% CL set on the gluon-gluon production cross
section of a SM-like Higgs boson as function of its mass.
Right: Similar for VBF production alone.
In both cases the blue line represents the theoretical expectation.}
\label{fig:abslimits_SM}
\end{figure}

\begin{figure}
\begin{minipage}{0.32\linewidth}
\centerline{\includegraphics[width=0.8\linewidth]{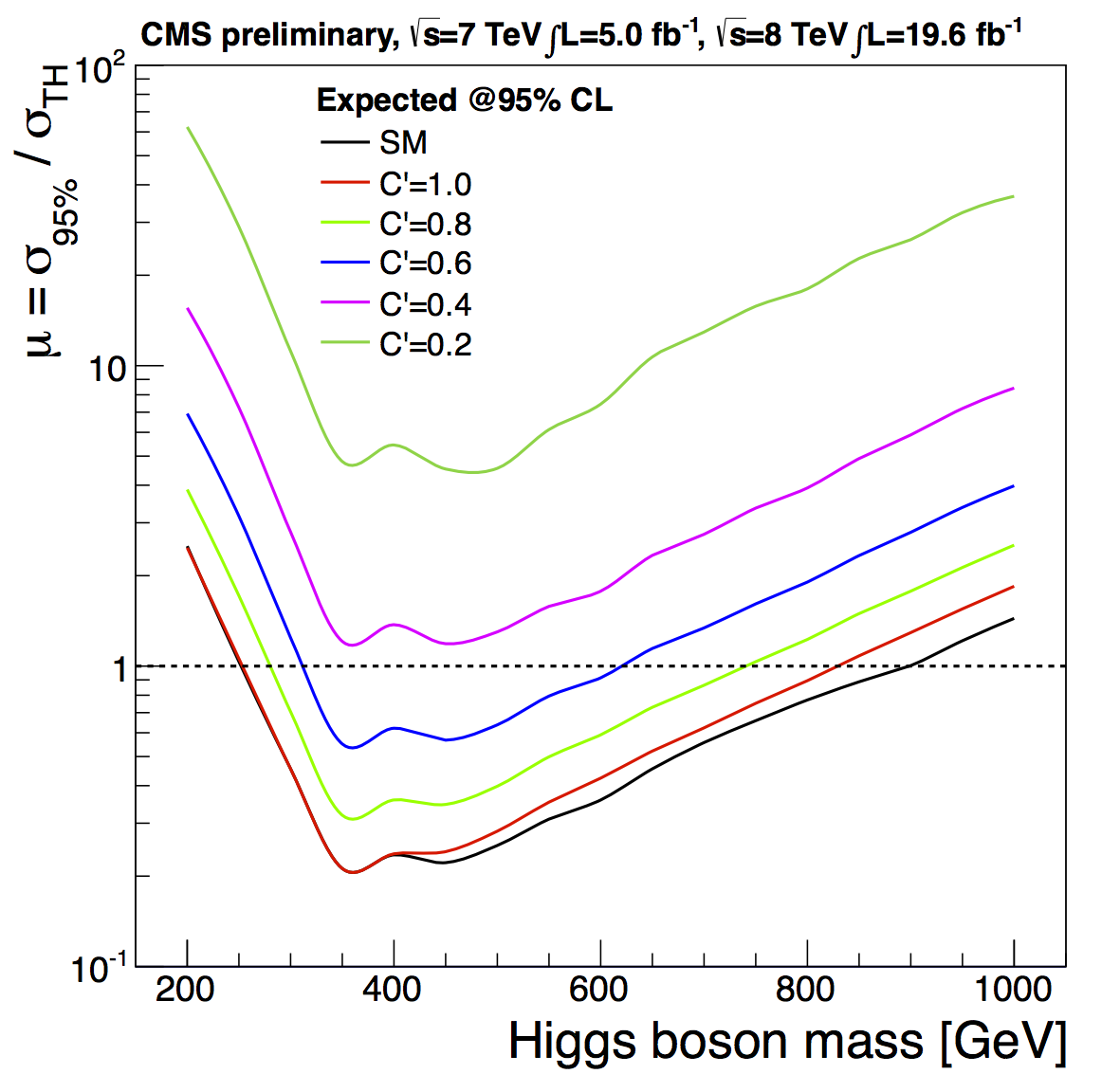}}
\end{minipage}
\hfill
\begin{minipage}{0.32\linewidth}
\centerline{\includegraphics[width=0.8\linewidth]{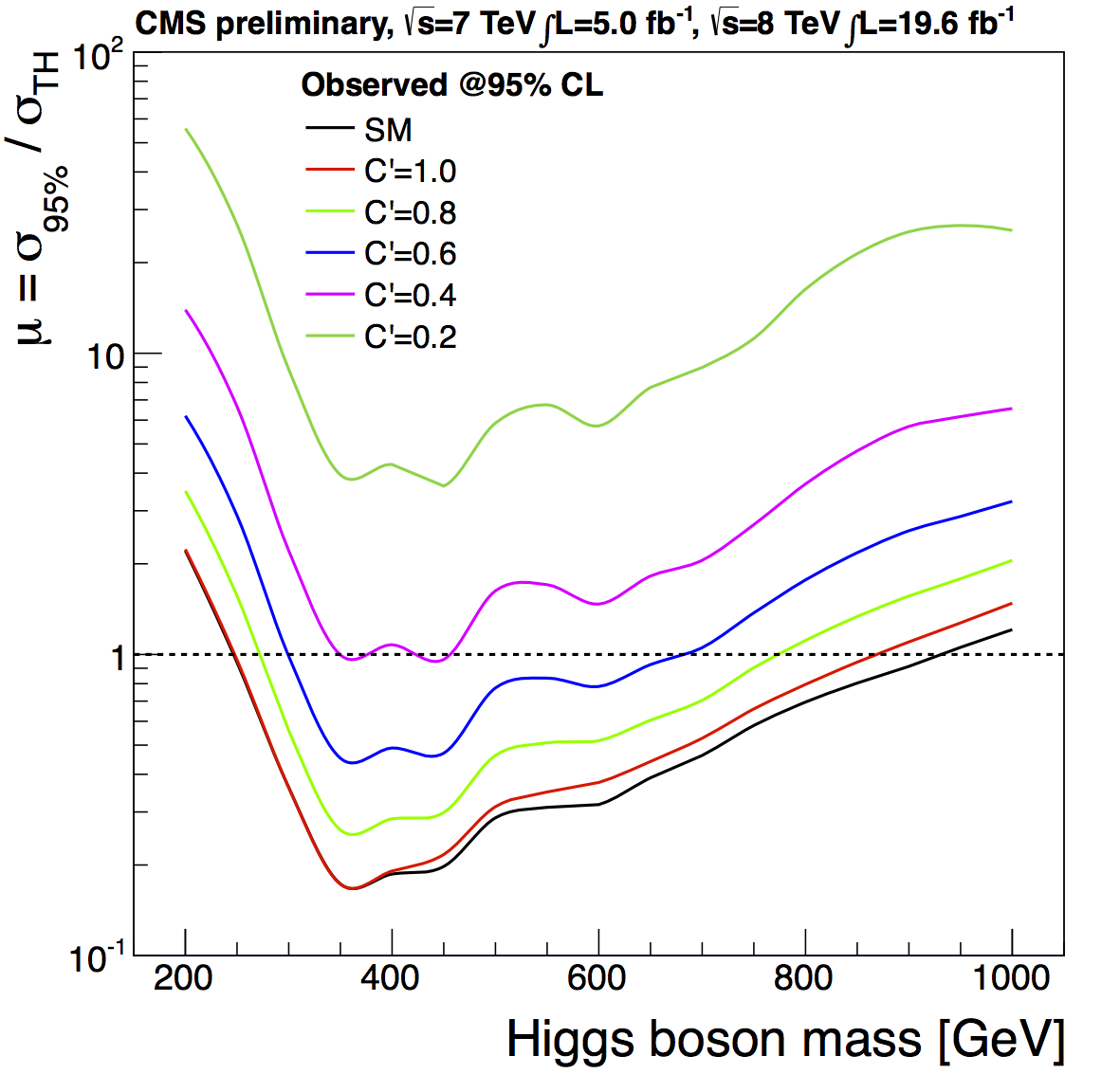}}
\end{minipage}
\hfill
\begin{minipage}{0.32\linewidth}
\centerline{\includegraphics[width=0.8\linewidth]{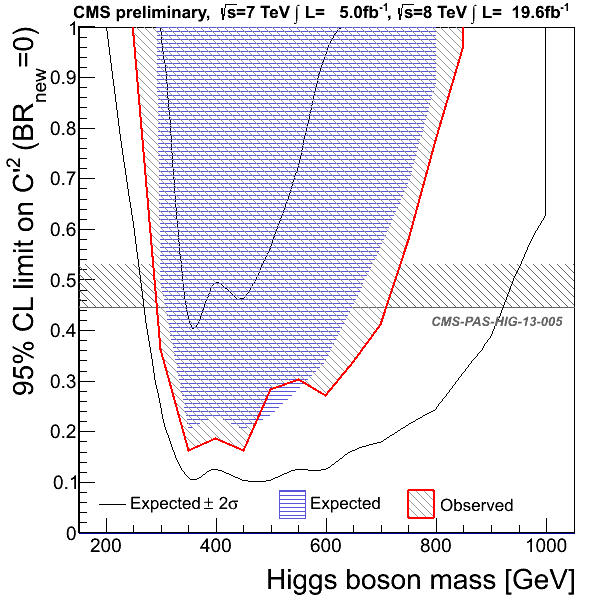}}
\end{minipage}
\caption{\textit{Left}: Expected limits on signal strength, $\mu$, of the EW singlet with modified couplings and width with respect to the SM, as function of its mass. \textit{Center}: Observed limits on $\mu$. \textit{Right}: Evolution of 95\% CL limit on the scale ($C'^2$) of the coupling of the EW singlet as function of its mass. To obtain these limits we have assumed that the new scalar does not decay to any new particles. In all scenarios the theory prediction is represented by the horizontal dashed line. Both data periods are combined.}
\label{fig:limits_narrow}
\end{figure}

Upper limits are set on production cross sections of Higgs boson for different \mh~hypotheses using a modified frequentist method often called as $CL_{s}$ method. Figure~\ref{fig:abslimits_SM} shows the results obtained from the shape-based analysis for a SM-like heavy Higgs boson using the 8 TeV dataset. The VBF production component is shown separately after subtracting the gluon-gluon fusion signal contribution from this category.
By using the signal strength  corresponding to a $CL_{\rm s}=0.05$, i.e. the ratio of the 95\% CL cross section upper limit ($\sigma$)
relative to the theoretical expectations ($\sigma_\mathrm{SM}$) denoted as $\mu^{\rm 95\%CL}=\sigma^{\rm 95\% CL}/\sigma_\mathrm{SM}$, we can further combine the results from the 8 TeV analysis with the re-analysed 7 TeV dataset. A SM-like Higgs boson is excluded in the mass range 248--930 GeV at 95\% CL using the shape-based analysis. The observed limit is in good agreement with the expected 254--898 GeV exclusion interval at 95\% CL. For the less sensitive cut-and-count analysis we obtain an observed exclusion of 268--756 GeV while 265--761 GeV is expected at 95\% CL.\\
We re-interpret the previous results as a search for an electroweak singlet
scalar mixing with the recently discovered candidate with a mass close to 125 GeV.
Phenomenologically the couplings of the two gauge eigenstates (SM and singlet)
become inter-related by unitarity and the original coupling strength of the 
light Higgs boson is therefore reduced with respect to the SM case.
If we define $C$ ($C'$) as the scale factor of the couplings of the low (high) mass
with respect to the SM, one can write $C^2+C^{'2}=1$ as the unitarity condition to be preserved.
The EWK singlet cross-section is also modified by a factor, $\mu'$,
and the modified width is $\Gamma'$; they are defined as:
\begin{equation}
\mu'=C'^2\cdot (1-BR_{\rm new})
\label{eq:bsmstrength}
\end{equation}
\begin{equation}
\Gamma'=\Gamma_{\rm SM} \cdot \frac{C'^2}{1-BR_{\rm new}}
\end{equation}
where $BR_{\rm new}$ is the branching ratio of the EWK singlet to
non-SM-like decay modes.
Indirectly we can set an upper limit at 95\% CL on  $C^{'2}<0.446$  using the signal strength fits to the H(125) candidate as obtained  
in~\cite{CMS-PAS-HIG-13-005}. In this analysis we have focused on the case where $C'^2\leq (1-BR_{\rm new})$. The SM signal mass line shape generated by $POWHEG$ is re-weighted to target a relativistic Breit-Wigner line shape with a narrower signal width with respect to the width of the SM Higgs boson. The contribution from the interference term between the BSM Higgs and the background is furthermore assumed to scale according to the modified coupling of the Higgs boson as: $(\mu+I)_{\rm BSM}=\mu_{\rm SM}C'^2+I_{\rm SM}C'$ where $\mu$ ($I$) is the signal strength (interference) in the BSM and SM cases. This assumption is based on the hypothesis that the couplings are similar to the SM case and simply re-scaled due to unitarity constraints. If we assume that the relative contribution from gluon-gluon fusion and VBF production modes of the new scalar is the same as the SM case we can derive the exclusion limits on the signal strength as a function of the Higgs boson mass summarized in Fig.~\ref{fig:limits_narrow}. We are able to combine 7 TeV and 8 TeV data under this assumption.
With this analysis one can also derive constraints on discovered boson decay width using its off-shell production. The observed (expected) exclusion limit at the 95\% CL is $\Gamma_{H}\leq6.4(10.7) \times \Gamma_{H}^{SM}$. The details of the analysis can be found in Ref.~\cite{14002}.

\end{document}